\documentclass[%
 reprint,
 amsmath,amssymb,
 aps,
]{revtex4-2}

\usepackage{graphicx}
\usepackage{dcolumn}
\usepackage{bm}
\usepackage{orcidlink}
\usepackage{hyperref}
\hypersetup{
    colorlinks=true,
    linkcolor=blue,
    urlcolor=black,
    citecolor=blue
    }
    \usepackage{titlesec}

\begin{document}

\preprint{APS/123-QED}

\title{Brillouin-Mandelstam scattering in telecommunications optical fiber at millikelvin temperatures}

\author{E. A. Cryer-Jenkins\,\orcidlink{0000-0003-2549-0280}\,$^{1,\dagger}$}
\author{A. C. Leung\,\orcidlink{0000-0001-9420-292X}\,$^{1,\dagger}$}
\author{H. Rathee\,\orcidlink{0009-0006-7762-8543}\,$^{1,\dagger}$}
\author{A. K. C. Tan\,$^{1,\dagger}$}
\author{K. D. Major\,\orcidlink{0000-0002-3268-6946}\,$^{1}$}
\author{M. R. Vanner\,\orcidlink{0000-0001-9816-5994}\,$^{1,}$}
\email{www.qmeas.net (m.vanner@imperial.ac.uk)}

\address{$^1$Quantum Measurement Lab, Blackett Laboratory, Imperial College London, London SW7 2BW, United Kingdom\\
$^{\dagger}$These authors contributed equally and have been listed alphabetically.
}

\begin{abstract}
Brillouin--Mandelstam scattering is a strong and readily accessible optical nonlinearity enabling a wide array of applications and research directions. For instance, the three-wave mixing process has been employed to great success for narrow-linewidth lasers, sensing applications, microscopy, and signal processing. While most of these avenues focus on room temperature operation, there is now increasing interest in cryogenic operation owing to the scattering mechanism's significant potential for applications and fundamental physics at low temperatures. Here, we measure the Brillouin scattering spectrum in standard single-mode telecommunications optical fiber at millikelvin temperatures using a closed-cycle dilution refrigerator and optical heterodyne detection. Our experiments are performed with a cryostat temperature from 50\,mK to 27\,K, extending previously reported measurements that utilized liquid helium-4 cryostats with temperatures greater than 1\,K. At millikelvin temperatures, our experiment observes coherent acoustic interaction with microscopic defects of the amorphous material---two-level-systems (TLS)---which has not been previously observed in optical fiber. The measured behaviour of the linewidth with temperature is in agreement with well-established models of ultrasonic attenuation in amorphous materials comprising a background intrinsic scattering, thermally-activated scattering, and incoherent and coherent TLS interaction. This work provides a foundation for a wide range of applications and further research including sensing applications, new approaches to investigate TLS physics, and Brillouin-scattering-based quantum science and technology.
\end{abstract}
\date{\today}

\maketitle


\section{\label{sec:intro}Introduction}
Brillouin-Mandelstam scattering is a three-wave-mixing process that couples two optical fields and a traveling acoustic wave. First theoretically described by Brillouin~\cite{Brillouin_1922} and Mandelstam~\cite{Mandelstam_1926} in the early 20th century, it has since grown into a rich research field~\cite{sipe2016hamiltonian, wiederhecker2019brillouin, wolff2021brillouin, merklein100YearsBrillouin2022} encompassing a broad spectrum of applications including low-threshold narrow-linewidth lasers~\cite{grudinin2009brillouin, otterstrom2018silicon, gundavarapu2019sub}, biological sensing~\cite{scarcelli2008confocal,prevedel2019brillouin}, microscopy~\cite{koski2005brillouin}, accelerometry~\cite{zarinetchi1991stimulated,li2017microresonator}, microwave filtration and synthesis~\cite{li2013microwave,pant2014chip,marpaung2015low}, and optomechanics applications~\cite{enzian2018observation, kim2019dynamic, kharel2022multimode, wang2024taming, cryer2023second}. Energy and momentum conservation allow for two scattering processes to occur: Stokes scattering, where the scattered light is down-shifted in frequency, and anti-Stokes scattering, where the scattered light is up-shifted in frequency. Driving various combinations of these two interactions enables the rich variety of applications highlighted above and Stokes scattering is used here for spectroscopy purposes.

Previous work cooling acoustic modes in Brillouin-scattering experiments has employed both cryogenic and laser-cooling-based techniques. Laser cooling is achieved by driving the anti-Stokes interaction, and experiments have cooled acoustic modes in resonators~\cite{bahl2012observation, renninger2018bulk} and photonic-crystal fibers~\cite{blazquez2024optoacoustic} with several experiments focusing on room-temperature initial conditions. By contrast, utilizing high acoustic frequencies in cryostats reaching less than 1\,K readily enables thermal occupations below unity to be achieved without laser cooling~\cite{Vacher1980, doeleman2023}.

Physical systems for Brillouin scattering utilize either amorphous or crystalline materials, each having their own advantages and disadvantages. Crystalline materials possess long-range order which enables low acoustic losses when at cryogenic temperatures~\cite{cole2010megahertz, galliou2013extremely, renninger2018bulk} while amorphous materials can be readily fabricated into various resonators and waveguides, with optical fibers being a prominent example. Such amorphous materials, owing to their irregular atomic arrangement~\cite{warren1937x}, display anomalous thermal and acoustic properties~\cite{zeller1971thermal, heinicke1971low, anderson1972anomalous} and microscopic defects---often referred to as two-level systems (TLS), which are theorized to be individual or small groups of atoms tunnelling between two energetically similar configurations---dominate the low-temperature acoustic properties of glasses. TLS exist over a very broad frequency range~\cite{hunklinger1973anomalous,Hunklinger1976,arcizet2009cryogenic,heinicke1971low, Vacher1980}, affect mechanical and electromagnetic oscillators, and are now a key challenge prevalent throughout many areas of cryogenic research and development including quantum optomechanics and superconducting circuits~\cite{muller2019towards}. In single-mode optical fiber, several works have begun to probe acoustic properties at cryogenic temperatures~\cite{lefloch2001experimental, leflochStudyBrillouinGain2003, thevenaz2002brillouin}. However, operation using liquid helium~4 limited temperatures to $\gtrsim$1\,K where many TLS are still thermally excited, suppressing coherent interactions between phonons and defects. As a result, the observation of coherent TLS interactions in optical fiber remained outstanding and it was not known what magnitude the effect would have in comparison to bulk silica, or even if the effect could be observed, owing to the fibre structure and germania dopants.

In this work, we measure the backward Brillouin scattering spectrum from 50\,mK to 27\,K in single-mode telecommunications optical fiber using a closed-cycle dilution refrigerator, 1548.0\,nm  optical pump pulses, and heterodyne detection of the weakly driven Stokes scattering signal. From the observed spectra, we extract the Brillouin frequency and linewidth, and their dependence on temperature displays the characteristic contributions from background intrinsic scattering, thermally activated scattering, and incoherent and coherent TLS interactions. These contributions and the associated material parameters are then estimated via numerical fitting of the linewidth and contrasted to the previous experimental literature. In particular, our measurements are consistent with previous measurements in optical fiber reported in Refs~\cite{lefloch2001experimental, leflochStudyBrillouinGain2003, thevenaz2002brillouin}, which characterized the Brillouin spectrum down to 1.2\,K. Notably, the lower-temperature measurements provided by our work enabled the characterisation of the coherent acoustic interaction with TLS, yet to be observed in optical fiber, which gives rise to an increase in the Brillouin linewidth at low temperatures. These observations cement optical fiber as a viable material platform to pursue TLS research with its ease-of-use and mature and tailorable fabrication processes. Moreover, they pave the way for a broad range of applications and further studies with prominent examples including extending fiber sensing to the millikelvin regime, further studies of TLS physics and its mitigation, and opening new avenues for quantum sensing and Brillouin scattering in the quantum regime~\cite{enzian2021single, enzian2021non, zhu2024optoacoustic}.

\section{\label{sec:method}Experimental method}
\subsection{\label{sec:setup}Experimental setup}
\begin{figure*}
\includegraphics[width=\textwidth]{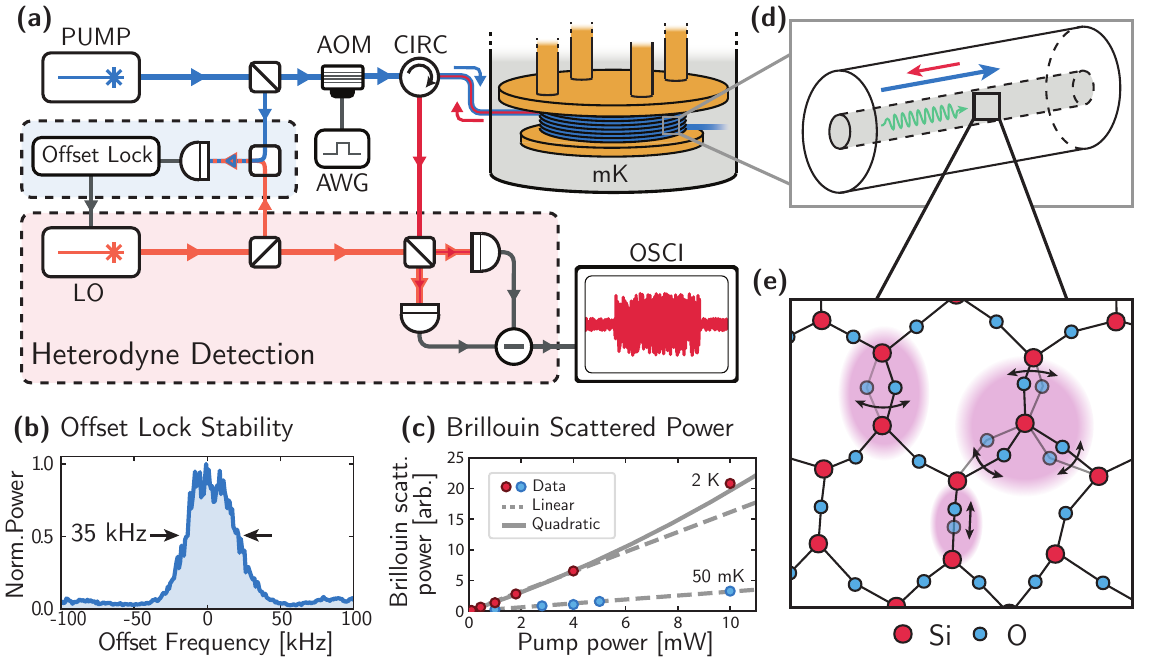}
\caption{\label{fig:setup} Scheme and experimental setup for Brillouin scattering in single-mode optical fiber at millkelvin temperatures. (a) Experimental schematic. A pump laser is pulsed using an acousto-optic modulator (AOM) driven by an arbitrary waveform generator (AWG) and coupled into a 50\,m length of single-mode fiber thermally anchored to the mixing-chamber plate of a dilution refrigerator. The backscattered Stokes light is separated using a circulator (CIRC) and mixed with a local oscillator (LO) locked at a fixed frequency offset from the pump. An oscilloscope (OSCI) then records heterodyne time traces used to obtain the power spectral density of the Brillouin Stokes signal. (b) Measured power spectrum of the beatnote between the pump and LO. The beatnote displays a FWHM of 35\,kHz which is much smaller than any linewidths measured in the experiment. (c) Stokes scattered power for different input pump powers at 50\,mK (blue points) and  2\,K (red points). The deviation from linearity for higher optical powers indicates stimulation of the Brillouin interaction or possible heating of the acoustic mode. For this work, a power of 1\,mW was used where the deviation from linearity was less than 2.5\% at 2\,K and less than 1\% at 50\,mK. The linear and quadratic curves at 50\,mK are indistiguishable over the range of powers presented in the plot. (d) Schematic of the optical and mechanical fields in single-mode fused silica optical fiber. The darker grey core guides both optical and acoustic modes. The pump wave (blue) scatters from a co-propagating acoustic field (green), generating counter-propagating Stokes light (red). (e) Schematic of candidate two-level systems (TLS) in amorphous silica. Several examples of possible two-level defects are highlighted in pink, including transverse and longitudinal positions of the Si-O-Si bond and the collective motion of a group of atoms.}
\end{figure*}

The experimental setup used to measure the Brillouin spectrum is shown in Fig.~\ref{fig:setup}(a). A 1548.0\,nm pump laser is sent through a 200\,MHz acousto-optic modulator driven by an arbitrary-waveform generator to produce rectangular optical pulses. The pump pulses then drive Stokes Brillouin scattering in a 50\,m length of standard acrylate-coated single-mode telecommunications fiber (SMF-28), which is wound to a high-purity copper bobbin thermally anchored to the mixing-chamber plate of a closed-cycle dilution refrigerator. The temperature of this plate is controlled via heaters from 50\,mK up to 27\,K with a feedback stability of 1 mK. The backscattered Brillouin signal returning from the fiber is separated with a circulator, measured via heterodyne detection, and we examine the $\sim11$\,GHz longitudinal acoustic mode with the lowest-order radial structure, which has maximal overlap with the optical pump mode. The local oscillator (LO) used for this measurement is generated by a second laser that is stabilised at a fixed frequency offset from the pump laser close to the Brillouin shift. The measured offset lock stability is shown in Fig.~\ref{fig:setup}(b) and had a width of 35\,kHz FWHM, far narrower than the Brillouin signals in this experiment. The heterodyne signal is recorded by an oscilloscope in the time-domain triggered by the arbitrary-waveform generator. For the lowest temperatures, $1.5\times10^{6}$ traces are recorded to obtain sufficient signal-to-noise (SNR) of the Brillouin spectra. Throughout the experiment, the power of the pump laser and LO are stabilised to ensure consistency across the full temperature range.

The pump pulse duration used was 5 $\mu$s, which is sufficiently long to resolve the smallest Brillouin linewidth in the experiment. The pulse repetition rate was 500 pulses s$^{-1}$ (a 0.25\% duty cycle), which is the maximum speed the oscilloscope can acquire while also ensuring no heating was observed within each pulse as well as over the measurement run. A peak power of 1\,mW during the pulse is used, which was chosen after measuring the Stokes-scattered power with input pump power at the lowest temperature of 50\,mK and at 2\,K where the narrowest Brillouin linewidth was observed, see Fig.~\ref{fig:setup}(c). At 50\,mK, a deviation from linearity is observed for pump peak powers exceeding 16\,mW, and, at 2\,K, the scattered power clearly deviates from a linear relationship above 5\,mW. We attribute this nonlinear response to a combination of stimulation of the Brillouin scattering process, and possible optically-induced heating~\cite{macdonald2016optomechanics} of the acoustic mode. Stimulation of the scattering process results from the Stokes interaction being proportional to the amount of scattered light and acoustic excitation, which themselves grow via the Stokes scattering. Optically-induced heating may occur if energy in the pump is converted into heat via absorption that increases the acoustic population to scatter from, which also increases the Stokes-scattered power. We have thus kept the pump peak power at 1\,mW for all of the measurements reported here, where the deviation from linearity is observed to be less than 2.5\%, to minimize perturbations to the acoustic state. It is also worth noting here that a third nonlinearity can exist where saturation of the TLS can reduce the acoustic decay rate, however, this effect is expected to be negligible for the pump powers used in this work~\cite{behunin2017engineering}. Further details of these three mechanisms, derived for a simple waveguide model, are discussed further in \hyperlink{nonlinear_gain}{Appendix C}.

A schematic of the geometry of optical single-mode fiber is shown in Fig.~\ref{fig:setup}(d). The SMF-28 fiber is made from high-purity fused silica with a (germanium dioxide) dopant to create a core of higher refractive index within a cladding. The higher dopant concentration also lowers the sound velocity of the core, simultaneously guiding the acoustic mode and providing an extended interaction length for Brillouin scattering to occur~\cite{jen1986leaky, yeniay2002spontaneous,kobyakov2005design}. Within the amorphous material, there are numerous possible microscopic defects in the arrangement of the atoms that give rise to TLS with a broad distribution of energies (cf. Fig.~\ref{fig:setup}(e)). At low temperatures, acoustic properties are predominantly governed by interactions with these defects, and their behaviour at different temperatures is discussed further in \hyperlink{sec:GlassPhysics}{Appendix A}. 

\subsection{\label{sec:analysis}Data analysis}
The power spectral density was computed from each of the triggered heterodyne time traces and averaged over the ensemble of measurements. The average power spectrum is then normalized to the vacuum power spectrum, which is obtained when there is no signal input to the heterodyne detector, and the spectrally-separated room-temperature component of the spectrum subtracted. The processed power spectra are then fitted with a Lorentzian function to extract the centre frequency and linewidth of the Brillouin spectrum. To estimate the temperature of the acoustic mode itself, we calibrate the scattered Stokes power at temperatures above 5\,K where good agreement is found between this work and previous studies in optical fiber~\cite{lefloch2001experimental, leflochStudyBrillouinGain2003} and utilise the theoretical temperature dependence of the scattered power to infer the mechanical mode temperature. Further details of this calibration procedure are given in \hyperlink{sec:mode_temp}{Appendix D}.

\section{\label{sec:results}Results}
\begin{figure*}
\centering
\includegraphics[width=0.8\textwidth]{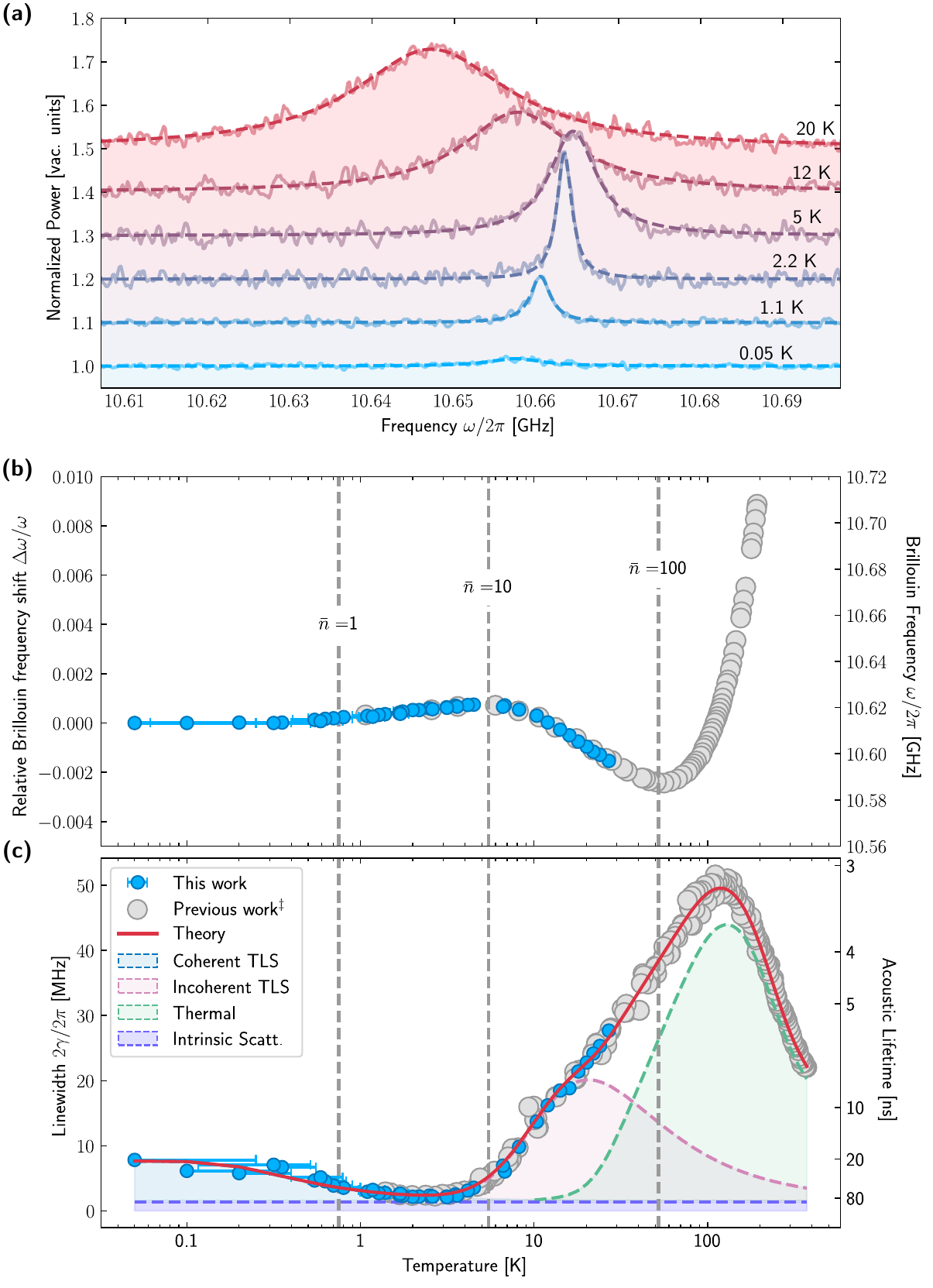}
\caption{Experimental characterization of Brillouin scattering at low temperatures. (a) Example measured power spectra at different temperatures. Each spectrum is normalised to the heterodyne vacuum response when no signal is present. Solid lines are experimental data and dashed curves are Lorentzian fits to extract linewidth and frequency. Each spectrum is offset by 0.1 vacuum units to aid visual comparison. (b) and (c) Plots of the measured relative Brillouin shift and linewidth, respectively, as a function of acoustic mode temperature with error bars indicating the temperature uncertainty. The corresponding absolute Brillouin shift and acoustic lifetime are also given. Light blue points are the experimental data points from this work, and the grey points are from previous work$^{\ddagger}$ performed above 1.4\,K. The difference in point size is for visual comparison only - it does not represent a difference in relative uncertainty between works. The red solid curve in (c) is a fit using the model described in \protect\hyperlink{sec:GlassPhysics}{Appendix A} and the shaded regions illustrate the contributions from the different damping mechanisms considered. Vertical dashed lines indicate mean occupations $\bar{n}$ as calculated from Bose--Einstein statistics. The onset of coherent TLS interactions is clearly associated with temperatures below where $\bar{n}=1$ where the relevant TLS have a high probability of being in their ground state. In both plots (b) and (c), offsets of -9.3\,MHz and +1.95\,MHz respectively have been applied to the data from previous work$^{\ddagger}$. Previous work$\ddagger$: Refs~\cite{lefloch2001experimental,leflochStudyBrillouinGain2003}.}
\label{fig:Fig2}
\end{figure*}

Fig.~\ref{fig:Fig2}(a) shows an example subset of the measured Brillouin power spectra from the full range of temperatures recorded. A clear non-monotonic behaviour is observed in both the Brillouin frequency and linewidth of the spectra with temperature due to the interplay of the different physical mechanisms affecting the acoustic field.

Fig.~\ref{fig:Fig2}(b) shows the measured relative Brillouin frequency shift $\Delta\omega/\omega\equiv(\omega(T)-\omega(T \rightarrow 0))/\omega(T \rightarrow 0)$ and Brillouin frequency $\omega/2\pi$ with temperature $T$. On the same figure, measurements made in previous works~\cite{lefloch2001experimental, leflochStudyBrillouinGain2003} from 1.4 to 370\,K are overlaid to illustrate the dependence over a larger temperature range and for comparative purposes. In the overlapping temperature range from 1.4 to 27\,K, our measurements are consistent with the previous works. Regarding the absolute Brillouin frequency, the data presented in the previous work~\cite{leflochStudyBrillouinGain2003} is offset by a  constant $-9.3$\,MHz compared to the measurements in this work. This small offset is likely due to the use of a different pump wavelength between our experiments and may also arise from differences in dopant levels, strain, and fiber geometry.

Fig.~\ref{fig:Fig2}(c) shows the Brillouin intensity linewidth FWHM $2\gamma/2\pi$ and associated lifetime ($1/2\gamma$) over the same temperature range overlaid with the same previous works~\cite{lefloch2001experimental, leflochStudyBrillouinGain2003}. Similarly to the frequency, we have applied an offset of +1.95\,MHz to the previous works' linewidth before plotting, and we attribute this difference to fiber geometry, dopant levels, inhomogeneous strain, and systematics from the larger laser linewidth used in Refs~\cite{lefloch2001experimental, leflochStudyBrillouinGain2003}. The Brillouin linewidth over 1.4 to 27\,K, where both datasets overlap, shows good qualititave agreement and the smallest observed acoustic linewidth in our experiment is 2.11\,MHz, corresponding to a phonon lifetime of 75.4\,ns. Notably, in our experiment, below 2\,K we observe a monotonic increase in the linewidth until the lowest temperature measured at 50\,mK. This increase is caused by the TLS having an increasing population in their ground state and therefore an enhanced ability to coherently absorb energy from the acoustic wave~\cite{Hunklinger1976}. The mean free path, and consequently the lifetime, of acoustic phonons is therefore reduced, resulting in an increasing Brillouin linewidth with decreasing temperature. An interesting side note is in Ref.~\cite{leflochStudyBrillouinGain2003} the onset of this phenomena was observed between 1.4 and 2\,K, however, the authors of the work remarked that the small increase in linewidth observed was likely due to analysis error. By extending the measured domain to millikelvin temperatures, our work reveals that the small increase in linewidth at $\gtrsim 1.4$\,K reported in Ref.~\cite{leflochStudyBrillouinGain2003} was, given the agreement between our data sets, more likely the onset of coherent TLS interaction.

\section{\label{sec:discussion}Discussion and Outlook}

To better understand the temperature dependence of the measured Brillouin spectra, the Brillouin linewidth is fitted with a model containing four mechanisms: (i) temperature-independent scattering, (ii) thermally-activated scattering, and (iii) incoherent and (iv) coherent TLS interaction. The fit (red solid line) and relative contribution of each mechanism (dashed lines) are shown in Fig.~\ref{fig:Fig2}(c). Firstly, a temperature-independent scattering rate is considered arising predominantly from acoustic Rayleigh scattering from fiber dopants and inhomogeneities. At temperatures above $\sim30$\,K, a thermally-activated relaxation process dominates the acoustic behaviour---the travelling acoustic wave modulates the energy distribution of an ensemble of structural defects in the fused silica, resulting in rethermalization and an irreversible exchange of heat with the environment that results in acoustic damping. At temperatures between 1 and 30\,K, a different relaxation process begins to dominate, involving the phonon-assisted tunnelling of TLS between their two states as the acoustic field again modulates their energy levels. For temperatures below $\sim$1\,K, where the thermal occupation is low, i.e. $\bar{n} \lesssim 1$, the ground state probability of the resonant TLS begins to approach unity. In this regime, the coherent absorption of phonons by TLS begins to dominate the observed behaviour. As the TLS ground-state population asymptotes to unity as the temperature decreases, acoustic phonons are more readily absorbed by the TLS, which can then decay via other channels leading to an increased damping of the acoustic field. This behaviour plateaus in the millikelvin range as illustrated by the lowest temperatures in Fig.~\ref{fig:Fig2}(c). Further details about the model and a table of the numerical fitting results are given in \hyperlink{sec:GlassPhysics}{Appendices A} and \hyperlink{sec:numerical_params}{B} respectively.

Utilising a closed-cycle dilution refigerator and heterodyne detection, we measure the backward Brillouin scattering spectrum from 50\,mK to 27\,K in a 50\,m length of single-mode telecommunications optical fiber. By calibrating the Stokes-scattering power observed with mixing-chamber-plate temperature, we infer the acoustic mode temperature to within a standard error of 201\,mK. Future experiments by our team will explore utilizing calibration-free sideband asymmetry measurements to reduce the uncertainty of this temperature. The Brillouin shift and linewidth observed are in good agreement with previous studies above 1\,K~\cite{lefloch2001experimental, thevenaz2002brillouin, leflochStudyBrillouinGain2003} and the variation in linewidth with temperature is well captured by current understanding of acoustic fields in amorphous materials. These findings extend previous work to the millikelvin regime enabling the first observation of coherent interactions between TLS and phonons in optical fiber. 

This work provides a foundation for a wide spectrum of further studies and new applications. For instance, this characterization of the Brillouin spectrum at low temperatures can enable the extension of optical fiber sensing~\cite{shimizuCoherentSelfheterodyneDetection1993, parkerTemperatureStrainDependence1997, maughan57kmSingleendedSpontaneous2001, bao22kmDistributedTemperature1993} to the millikelvin regime, providing a promising and inexpensive new tool for millikelvin cryogenic applications that can also achieve high spatial resolution~\cite{songDistributedStrainMeasurement2006}. Our results also help to demonstrate that optical fibers provide an accessible test bed for further studies of the physics of TLS which are remarkably universal in amorphous materials affecting all electromagnetic and mechanical oscillators. Indeed, the mitigation of TLS-induced damping is of immediate importance to quantum optomechanics and superconducting quantum circuits~\cite{muller2019towards}. One technique to reduce the damping contribution from TLS is to perform spectral hole burning in the frequency band required~\cite{behunin2017engineering}, which can be explored here to reduce the coherent TLS contribution at millikelvin temperatures. The characterization of the Brillouin spectrum at millikelvin temperatures and the experimental techniques developed here also provides a foundation to explore Brillouin scattering in the quantum regime.

\section*{Funding Information}
A.K.C.T. acknowledges support from A*STAR Singapore through the National Science Scholarship. This project was supported by UK Research and Innovation (MR/S032924/1, MR/X024105/1), the Engineering and Physical Sciences Research Council (EP/T031271/1), and the Science and Technology Facilities Council (ST/W006553/1).

\section*{Acknowledgments}
We acknowledge useful discussions with Georg Enzian, Lars Freisem, Niall Moroney, John J. Price, Andreas {\O}. Svela, and Magdalena Szczykulska. 


\bibliography{bibliography}


\onecolumngrid
\clearpage
\setcounter{equation}{0}
\setcounter{section}{0}
\setcounter{subsection}{0}
\setcounter{subsubsection}{0}
\def\theequation{\Alph{section}\arabic{equation}}
\counterwithin*{equation}{section}
\def\thefigure{\arabic{figure}}
\def\thetable{\Roman{table}}
\def\thesection{\Alph{section}}
\def\thesubsection{\arabic{subsection}}
\def\thesubsubsection{\alph{subsubsection}}

\titleformat{\section}
  {\normalfont\bfseries\centering} 
  {Appendix \thesection:}        
  {0.5em}                        
  {}                           

\begin{center}
\textbf{\large{Brillouin-Mandelstam scattering\\in telecommunications optical fiber at millikelvin temperatures\\Appendices}}
\end{center}
\vspace{10pt}

\begin{centering}
E. A. Cryer-Jenkins\,\orcidlink{0000-0003-2549-0280}\,$^{1,\dagger}$,
A. C. Leung\,\orcidlink{0000-0001-9420-292X}\,$^{1,\dagger}$,
H. Rathee\,\orcidlink{0009-0006-7762-8543}\,$^{1,\dagger}$,
A. K. C. Tan\,$^{1,\dagger}$,
K. D. Major\,\orcidlink{0000-0002-3268-6946}\,$^{1}$, and
M. R. Vanner\,\orcidlink{0000-0001-9816-5994}\,$^{1,*}$
\end{centering}

\vspace{16pt}

\begin{centering}
\textit{\small
$^1${Quantum Measurement Lab, Blackett Laboratory, Imperial College London, London SW7 2BW, United Kingdom}\\
$^{\dagger}$These authors contributed equally to this work and are listed alphabetically.\\
$^{*}${www.qmeas.net (m.vanner@imperial.ac.uk)}\\
}
\end{centering}

\section{Acoustic Fields in Amorphous Glasses}
\hypertarget{sec:GlassPhysics}{}
In this Appendix, we first describe further details of the four damping mechanisms considered: (i) coherent TLS, (ii) incoherent TLS, (iii) thermally activated, and (iv) intrinsic scattering. For mechanisms (i) -- (iii), we discuss and summarize the models in Ref.~\cite{Hunklinger1976} by Hunklinger and Arnold. For mechanism (iv), we qualitatively list possible sources of temperature-independent damping.

\subsection{Two-Level Systems}
The thermal and acoustic properties of amorphous glasses display markedly different behaviours to those of crystalline materials that possess long-range order in constituent atom arrangement. These deviations are suggested to originate from excited states comprised of different structural configurations, intrinsic to the disordered nature of the glass itself. Often referred to as two-level systems (TLS), the relatively low spatial density of these defects make them difficult to analyse computationally although several proposals exist for what physical configurations may comprise them informed by the characteristic energy of the TLS. Three potential candidates for fused silica are illustrated in Figure~\ref{fig:setup}(e) of the main text and demonstrate how the disordered nature of the glass renders each defect unique, resulting in a broad array of decay channels for acoustic waves. It is through the interaction with these localised two-state defects that the variation in Brillouin shift and linewidth at low temperatures is understood to arise, which contribute to the dispersive and dissipative behaviour of acoustic waves through two main mechanisms---coherent interaction, and incoherent relaxation.

\subsubsection{Coherent TLS}
\label{subsection:resonant_tls}
\begin{figure}
    \centering
    \includegraphics{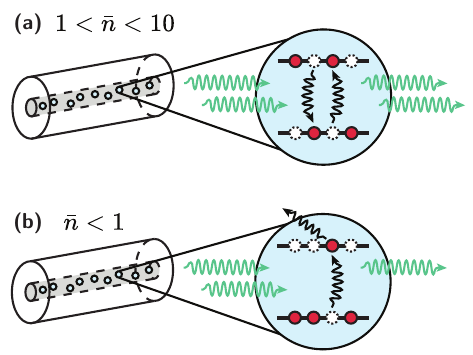}
    \caption{Coherent acoustic interactions with TLS. \textbf{(a)} At temperatures above a thermal occupation of unity, the TLS have significant population in both the ground and excited states and coherent absorption and stimulated phonon emission can occur with similar rates. \textbf{(b)} At lower temperatures where the thermal occupation is below unity, only coherent absorption can take place, which can then decay via multiple channels, giving rise to damping of the acoustic mode.}
    \label{fig:ResonantTLS}
\end{figure}

At very low temperatures ($k_B T/\hbar\omega \lesssim 1$), the ground state probability of the TLS with energy spacing matching the acoustic frequency begins to approach unity and coherent absorption of phonons, followed by decay via other channels, is the dominant scattering mechanism, see Fig.~\ref{fig:ResonantTLS}. As such, an increase in acoustic absorption is observed as the temperature decreases. For the coherent process, the acoustic intensity decay rate $2\gamma_{C}/2\pi$ is given by~\cite{Hunklinger1976} 
\begin{equation}
    \frac{2\gamma_{C}}{2\pi} = \frac{n(E) M_l^2 E}{2\hbar \rho v_l^2} \text{tanh}(E/2 k_B T).
    \label{eq:resonant_linewidth}
\end{equation}
where $E = \hbar \omega$, $n(E)=n_0(1+a(E/k_B)^2)$ is the TLS density of states which is assumed to be weakly quadratic with  coefficient $a$ and zero-energy density $n_0$. $M_l$ is the deformation potential associated with the coupling between the longitudinal acoustic wave and the TLS, $\rho$ is the material density and $v_l$ is the speed of sound for longitudinal acoustic waves. As this coherent TLS interaction contribution to the acoustic linewidth depends on the phonon energy, and hence acoustic frequency, lower acoustic frequencies at low temperatures will have a lower decay rate. It should be noted that Eq.~(\ref{eq:resonant_linewidth}) is derived for the case that the acoustic intensity remains low in order that coherent absorption processes are not saturated which occurs when a significant portion of TLS are driven into their excited state~\cite{Behunin2016_longlived}.

\subsubsection{Incoherent TLS}
\label{subsection:off_resonant_TLS}
At slightly higher temperatures ($10 \lesssim k_B T/\hbar\omega \lesssim 100$), acoustic dissipation is instead dominated by a different process. A travelling acoustic wave modulates the energy splitting of TLS in its path, temporarily shifting them away from thermal equilibrium with their environment. The subsequent rethermalization is associated with an irreversible exchange of heat with the thermal bath, leading to a damping of the acoustic wave. As the TLS participating in this damping are not limited to those at the acoustic frequency and involves the exchange of thermal phonons of all polarizations~\cite{Hunklinger1976}, this mechanism is sometimes referred to as incoherent damping. The Brillouin linewidth $2\gamma_{I}/2\pi$ due to this mechanism is given by 
\begin{equation}
    \frac{2\gamma_{I}}{2\pi} = \frac{D_{I}^2}{2\pi\rho v_l^2} \int_0^{\infty} \! dE \, \frac{\exp(E/k_B T)}{k_B T[1 + \exp(E/k_B T)]^2} \frac{n(E) \omega^2 \tau_{I}(E)}{1 + \omega^2 \tau_{I}(E)^2} .
    \label{eq:off-resonant_linewidth}
\end{equation}
Here, $D_{I}$ is the deformation potential associated with the TLS energy shift induced by an acoustic wave for unit strain. The inverse relaxation time $\tau_{I}^{-1}$ is given by 
\begin{equation}
    \tau_{I}^{-1} = \left(\frac{M_l^2}{v_l^5} +\frac{2M_t^2}{v_t^5}\right)\frac{E^3}{2 \pi \rho \hbar^4} \text{coth}\left(\frac{E}{2 k_B T}\right) ,
    \label{eq:inverse_telax_time}
\end{equation}
where the $t$ subscript denotes transverse instead of longitudinal quantities.

As this work focuses on the quasi-longitudinal acoustic waves participating in backward Brillouin scattering, coherent interactions between TLS and transverse acoustic waves (parametrized by the analogous transverse deformation potential $M_t$) are not directly probed. However, the incoherent TLS contribution to the acoustic linewidth involves the coupling of defects to both longitudinal and transverse acoustic modes allowing both deformation potentials to be estimated, as is seen in Eq.~(\ref{eq:inverse_telax_time}). As the ratio of deformation energy density to elastic energy density for each acoustic polarisation obtained from the numerical fitting of our experimental results is approximately equal $n_0M_l^2/\rho v_l^2\approx n_0M_t^2/\rho v_t^2$, coupling between transverse acoustic waves and TLS in amorphous silica is then a similar magnitude to their longitudinal counterparts. This result is consistent with previous estimates of TLS-coupling to different types of acoustic wave experimentally explored in Refs~\cite{hunklinger1975propagation,Vacher1980}, and is supported by the fitted parameters from our experiment presented in Table~\ref{tab:fitted_params}.

\subsection{Thermally Activated Processes}
\label{subsection:thermally_activated_processes}
Due to the varied nature of the defects found in amorphous glasses and their lack of consistent energy splitting, relaxation processes can occur at several energy scales. For higher temperatures, a similar model proposed by Anderson and B\"ommel in Ref.~\cite{Anderson1955} considers two-state defects as particles in double-well potentials with a broad range of activation energies well described as another relaxation-mediated process~\cite{Anderson1955,leflochStudyBrillouinGain2003}. The contribution from this mechanism to the linewidth of the acoustic wave $2\gamma_{TA}/2\pi$ is given by 

\begin{equation}
    \frac{2\gamma_{TA}}{2\pi} = \frac{D_{TA}^2}{8\pi\rho v_{l}^2 k_B T} \int^{\infty}_{0} \! dE \, P(E) \frac{\omega^2 \tau_{TA}(E)}{1 + \omega^2 \tau_{TA}(E)^2} .
    \label{eq:thermally_activated_linewidth}
\end{equation}
Here, $D_{TA}$ is a different but analogous deformation potential to Eqn~(\ref{eq:off-resonant_linewidth}), $P(E)$ describes a distribution of activation energies for the excitation of the TLS, and $\tau_{TA}$ is the time for the relaxation process to reestablish equilibrium, dependent on the activation energy. For thermally activated transitions, this time is given by the Arrhenian relationship 
\begin{equation}
    \tau_{TA}(E) = \tau_0 \exp (E/k_B T) ,
\end{equation}
where $\tau_0$ is the characteristic time scale of the process and we follow Ref.~\cite{Hunklinger1976} in assuming a Gaussian distribution of energies with a low energy cutoff described by
\begin{equation}
    P(E) = n_{TA} \exp\left[-\frac{(E-E_0)^2}{2E_1^2} - \left(\frac{E_2}{E}\right)^2\right] ,
\end{equation}
where $n_{TA}$ is a normalization factor, $E_0$ and $E_1$ are the mean and standard deviation of the activation energy respectively, and $E_2$ is a phenomenological cutoff imposed to provide better agreement at low temperatures/high frequencies~\cite{Hunklinger1974}. The relaxation physics underpinning both the incoherent and thermally-activated contributions to the acoustic behaviour is manifest in the functional form of the integrands in Sections \ref{subsection:off_resonant_TLS} and \ref{subsection:thermally_activated_processes}. At these higher temperatures, another damping process resulting from the anharmonicity of acoustic coupling to the thermal bath~\cite{vacher2005anharmonic} may contribute but we do not currently consider this form of damping as we focus on the low-temperature range relevant for our measurements.

\subsection{Intrinsic Scattering}
Finally, we consider the temperature-independent intrinsic scattering $2\gamma_{IS}/2\pi$ present in the Brillouin linewidth to be the result of several mechanisms including acoustic Rayleigh scattering from inhomogeneities in the fiber such as the germanium dopants and other impurities, as well as a small inherent loss due to radiative damping as the fundamental acoustic wave couples to other unguided mechanical modes of the fiber mediated by the core-cladding boundary. Previous works~\cite{Behunin2016_longlived,jen1986leaky} have estimated the relative contributions of these temperature-independent mechanisms for similar fibers and suggest that dopant-related scattering is likely to contribute the majority of the intrinsic linewidth.

The measured intensity linewidth $2\gamma/2\pi$ of the acoustic wave at each temperature is then described by the sum of all the contributions discussed above
\begin{equation}
    \frac{2\gamma}{2\pi} = \frac{2(\gamma_{C} + \gamma_{I} + \gamma_{TA} + \gamma_{IS})}{2\pi} .
\end{equation}

\section{Numerical Fitting and Parameters}
\hypertarget{sec:numerical_params}{}
Numerical fitting using the model described in the previous section was carried out using a non-linear least squares algorithm and numerical integration of the equations capturing incoherent and thermally activated interactions. The inferred acoustic mode temperature is used for numerical fitting and the error bars in the inferred temperatures are utilised to determine the uncertainty in  fitted parameters. To reduce computation time, a finite upper integration limit was used that was tested for convergence at all temperatures involved in the fitted data.  The parameters obtained from the numerical fits and their respective uncertainties are given in Table~\ref{tab:fitted_params}. 

\addtolength{\tabcolsep}{+6pt} 
\begin{table*}[ht]
    \caption{Fitted parameters from the model described in \protect\hyperlink{sec:GlassPhysics}{Appendix A} with their uncertainties and their comparison with values from experimental literature on bulk vitreous silica. The standard errors in the thermal defect timescale $\tau_0$ and intrinsic scattering linewidth $2\gamma_{IS}/2\pi$ are much smaller than those of other quantities and are therefore not shown.}
    \centering
    \begin{tabular}{l c c c c}
        \hline\hline
         \textbf{Parameter} & \textbf{Symbol} & \textbf{Fitted Value} & \textbf{Literature Value} & \textbf{Units}\\ [0.5ex] 
         \hline
          Longitudinal Energy Density & $n_0M_l^2$ & $1.42(1)\times10^{7}$ &  $2.2\times10^{7}$ in Ref~\cite{Vacher1980} & Jm$^{-3}$\\
          Transverse Energy Density & $n_0M_t^2$ & $0.584(2)\times10^{7}$ & $1.3\times10^{7}$ in Ref~\cite{Vacher1980} & Jm$^{-3}$\\
          Incoherent Energy Density & $n_0D_{I}^2$ & $2.23(1)\times10^8$ & $2.6\times10^8$ in Ref~\cite{Vacher1980} & Jm$^{-3}$\\
          Curvature of density of states & $a$ & $5.16(1)\times10^{-3}$ & $3.5\times10^{-3}$ in Ref~\cite{Vacher1980} & K$^{-2}$ \\
          Mean of $P(E)$ & $E_0$ & 423(4) & 410 in Ref~\cite{Hunklinger1974} & K \\
          Width of $P(E)$ & $E_1$ & 580(10) & 550 in Ref~\cite{Hunklinger1974} & K \\
          Cut-off of $P(E)$ & $E_2$ & 130(10) & 125 in Ref~\cite{Hunklinger1974} & K \\
          Timescale of thermal defects & $\tau_0$ & $5.15\times10^{-13}$ & $1\times10^{-13}$ in Ref~\cite{Hunklinger1974} & s\\
          Intrinsic scattering linewidth & $2\gamma_{IS}/2\pi$ & 1.70 & 2.5 in Ref~\cite{Vacher1980} & MHz \\
          TLS density of states & $n_0$ & - & $7.8\times10^{45}$ in Ref~\cite{Stephens1973} & J$^{-1}$m$^{-3}$ \\
        \hline\hline
    \end{tabular}
    \label{tab:fitted_params}
\end{table*}
\addtolength{\tabcolsep}{-6pt} 

\section{Stokes-Scattering Pump-Power Dependence}
\hypertarget{nonlinear_gain}{}In this section we derive the Stokes-scattered power as a function of input pump power starting from the waveguide equations of motion for the system.

The Brillouin frequency and linewidth may be modified by the Brillouin interaction itself, optically-induced heating, and saturation of TLS-induced damping~\cite{behunin2017engineering}. In order to ensure that the damping observed is primarily due to unperturbed material properties, in our measurements we use a low pump power where the response of the Stokes-scattered power with pump power is linear. Here, we also illustrate that this linearity demonstrates that the unwanted effects of Brillouin stimulation, heating, and TLS saturation are negligible.

In a Brillouin-active waveguide such as an optical fiber, the equations of motion for the pump, Stokes, and mechanical mode envelope operators are given by~\cite{sipe2016hamiltonian,van2016unifying,zoubi2016optomechanical}:
\begin{align}
    &v_P^{-1}\partial_t a_P + \partial_z a_P = -ig_0a_Sb - (\alpha_P -i\Delta_P)a_P , \notag \\
    &v_S^{-1}\partial_t a_S - \partial_z a_S = -ig_0b^{\dagger}a_P - (\alpha_S -i\Delta_S)a_S , \\
    &v_M^{-1}\partial_t b + \partial_z b = -ig_0a_S^{\dagger}a_P - (\alpha_M -i\Delta_M)b + \sqrt{2\alpha_M}b_{in}, \notag
\end{align}
where $a_P(z,t)$, $a_S(z,t)$, and $b(z,t)$ are mode envelope annihilation operators, $v_{P/S/M}$ are group velocities, $\alpha_{P/S/M}$ are amplitude losses per unit length, $\Delta_{P/S/M}$ are wavevector offsets from the applied fields to the intrinsic waveguide modes, $\partial_{t/z}$ are partial derivatives with respect to time/space, $g_0$ is the traveling-wave three-wave-mixing coupling rate, and $b_{in}(z,t)$ is a thermal noise term with correlation function $\langle b_{in}(z,t)b^{\dagger}_{in}(z',t')\rangle=(\Bar{n}+1)\delta(t-t')\delta(z-z')$. Here, $\Bar{n}=1/(e^{\hbar\omega_B/k_BT}-1)$ is the mechanical occupation of the thermal bath associated with $b_{in}$ with Brillouin frequency $\omega_B$ and temperature $T$, and $\delta(z)$ is the Dirac $\delta$-function.

To obtain the backscattered Stokes power density as a function of mechanical wavevector $P_S(z;\Delta_M) = \hbar\omega_S\langle a^{\dagger}_Sa_S\rangle(z)$ (where $\omega_S$ is the Stokes angular frequency), we consider the system in its steady state $(\partial_t\rightarrow0)$ and assume an undepleted pump field $(\partial_z\langle a^{\dagger}_Pa_P\rangle\rightarrow0)$. We lastly assume that the acoustic field is strongly damped $(\alpha_S\ll\alpha_M)$ and therefore a localized quantity $(\partial_z b\rightarrow0)$ and that the input Stokes field is initially vacuum $(P_S(L;\Delta_M)=0)$. Following the approach outlined in Refs~\cite{van2016unifying,zoubi2016optomechanical}, one obtains for the backscattered Stokes power density at the optical fiber's input (i.e. at $z=0$)
\begin{equation}
    P_{S}(0;\Delta_M) = \frac{\hbar\omega_Sv_M(\bar{n}+1)G_B(\Delta_M) P }{\alpha_S-G_B(\Delta_M) P}\left(1-e^{-2[\alpha_S-G_B(\Delta_M) P]L}\right) ,
    \label{eq:StokesPower}
\end{equation}
where $L$ is the length of the optical fiber, $P=\hbar\omega_P\langle a^{\dagger}_Pa_P\rangle$ is the input optical pump power at pump angular frequency $\omega_P$, and
\begin{equation*}
    G_B(\Delta_M)=\frac{g_0^2}{\hbar\omega_P\alpha_M\left(1+\frac{\Delta_M^2}{\alpha_M^2}\right)} = \frac{\Bar{G_B}}{\left(1+\frac{\Delta_M^2}{\alpha_M^2}\right)} ,
\end{equation*}
is the wavevector-dependent Brillouin gain coefficient with on-resonance gain factor $\Bar{G_B}=g_0^2/\hbar\omega_P\alpha_M$.

Assuming that optical losses are negligible ($\alpha_S\rightarrow0$) and $P$ remains relatively small ($P\ll1/2G_B(\Delta_M)L$), one may expand the exponential term in Equation~(\ref{eq:StokesPower}) up to order $P^2$ and, integrating over $\Delta_M$, we obtain the total scattered Stokes power
\begin{equation}
    P_{S}(0) = 2\pi\hbar\omega_Sv_M\alpha_M(\bar{n}+1)\Bar{G_B}L\left(P+\frac{\bar{G_B}L}{2}P^2\right) .
    \label{eq:StokesPowerQuadratic}
\end{equation}
Equation~\ref{eq:StokesPowerQuadratic} contains terms both linear and quadratic in $P$, the former arising from thermal and spontaneous Brillouin scattering and the latter from the stimulation of the scattering process. Furthermore, assuming a simple linear dependence of the thermal occupation on pump power to account for optically-induced heating (i.e. $\Bar{n}(P)=\Bar{n}_0 + \zeta P$ where $\Bar{n}_0$ is the unheated bath occupation and $\zeta$ is a constant of proportionality), we can insert the latter expression into Equation~(\ref{eq:StokesPowerQuadratic}) and, again keeping terms to order $P^2$, obtain
\begin{equation}
    P_{S}(0) = 2\pi\hbar\omega_Sv_M\alpha_M(\bar{n}_0+1)\Bar{G_B}L\left(P+\left[\frac{\zeta}{\Bar{n}_0+1}+\frac{\bar{G_B}L}{2}\right]P^2\right).
\end{equation}
A similar procedure may be carried out for a power-dependent acoustic decay rate $\alpha_M(P)$ arising from the saturation of TLS-induced damping for large pump (and accordingly acoustic) powers which drive a large portion of TLS into their excited state where they can no longer absorb energy. Modeling the acoustic loss as $\alpha_M(P)=\alpha_{M,0} -\xi P$ yields a Stokes scattered power that is at least cubic in input pump power. We therefore observe that stimulated Brillouin scattering, optically-induced heating, and TLS saturation all lead to a nonlinear growth in the backscattered Stokes power for increasing pump power. Through measurements over a range of input powers and temperatures illustrated in Figure~\ref{fig:setup}(c), we are able to observe the deviation from a linear relationship and operate with an accordingly low power where the deviation from linearity is $<2.5\%$ for all temperatures.

Furthermore, the Stokes-scattered power (integrated over a 500\,ns window) was computed as a function of time within each 5\,\textmu s pulse and no appreciable heating dynamics were observed within the pulse duration. A longer timescale measurement of Stokes power, made over $\sim$400 minutes at 50\,mK, also did not display heating.

\section{Mode Temperature Estimation}
\begin{figure}
\includegraphics[width=0.4\columnwidth]{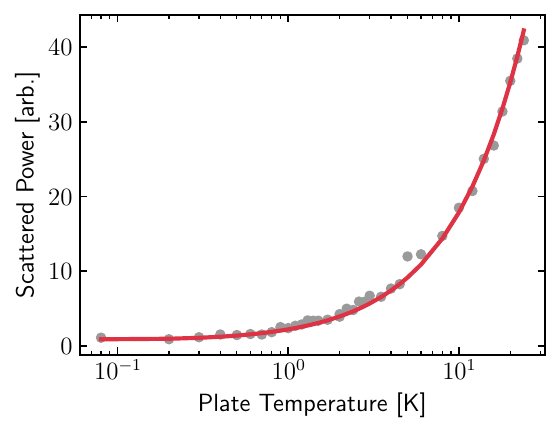}
\caption{\label{fig:stokes_area}Brillouin Stokes scattered power as a function of mixing-chamber-plate temperature. The red line is a fit of Eq.~\ref{eq:mode_temp} using points $>5$\,K and is dependent on the thermal acoustic population at different temperatures.}
\label{Fig:AppAreas}
\end{figure}

\hypertarget{sec:mode_temp}{}
To ensure that the Brillouin scattering characterized in this work is accurate at low temperatures, it is important to quantify how closely the mixing-chamber plate temperature relates to the the actual temperature of the acoustic mode.  To do so, we utilize the measured Stokes scattered power at each temperature to infer the mode temperature.  The measured Stokes scattered power $P_{Smeas}$ can be derived using Equation~\ref{eq:StokesPower} and, along with the definition of the mechanical occupation $\Bar{n}$, can be expressed as
\begin{equation}
    P_{Smeas}=A\left(\frac{1}{e^{\hbar\omega_B/k_BT}-1}+1\right) ,
    \label{eq:mode_temp}
\end{equation}
where $\omega_B/2\pi \approx 10.6$\,GHz is the Brillouin frequency in silica at low temperatures, and $A$ is a calibration factor that accounts for the system parameters and detection system gain and efficiency. Given that the pump pulse parameters and detection system are unchanged (and assuming negligible nonlinear gain as addressed in \hyperlink{nonlinear_gain}{Appendix C}), the value of $A$ is independent of temperature. Therefore, $A$ can be determined using the scattered Stokes power measured at temperatures $>5$ K where there is high confidence that plate temperatures well approximates the mode temperature. This is because at higher temperatures, possible optical heating from the measurements have a smaller impact and the measured Brillouin linewidth and frequency of this work are in good agreement with existing results~\cite{leflochStudyBrillouinGain2003}.  The Stokes scattered power for different temperatures is shown in Figure~\ref{fig:stokes_area}.

\begin{figure}
\includegraphics[width=0.4\columnwidth]{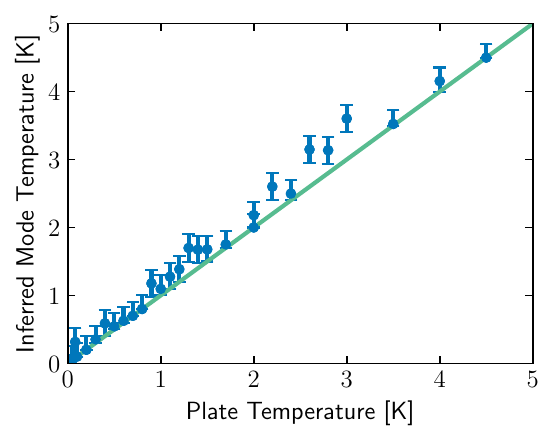}
\caption{\label{fig:thermometry}Inferred acoustic mode temperature with mixing-chamber-plate temperature obtained from the calibrated Stokes scattered power plotted in Fig.~\ref{Fig:AppAreas}. The green line indicates where the inferred mode temperature equals the mixing-chamber-plate temperature and, to be conservative, we do not allow the mode temperature and associated uncertainty to be below this line.}
\end{figure}

Using Equation~\ref{eq:mode_temp} to fit the $>5$\,K data in Figure \ref{fig:stokes_area} allows the Stokes scattered power for temperatures below 5\,K to be used to infer the actual mode temperature. In instances where the inferred temperature is lower than the plate temperature, the inferred temperature is raised to the plate temperature to be conservative as it is unlikely that the mode would be colder than the mixing-chamber-plate RuOx temperature sensor. The resulting inferred mode temperature is plotted against the plate temperature of the system in Figure~\ref{fig:thermometry}. From the fit and the scatter of experimental points, we obtain a standard error for the inferred acoustic mode temperature of 201\,mK. We also assume, to be conservative, that the error on each inferred temperature cannot extend below the mixing-chamber plate temperature for the same reasoning as outlined above.

\section{Experimental Parameters}
\hypertarget{sec:exp_params}{}
The parameters for our fiber experiment are listed in Table ~\ref{tab:exp_params}.

\begin{table}[ht]
\caption{Experimental parameters}
\centering 
\begin{tabular}{l c c c} 
\hline\hline 
\textbf{Parameter} & \textbf{Symbol} & \textbf{Value} & \textbf{Units} \\ [0.5ex] 
\hline 
Fiber length & & 50 & m \\
Fiber temperature & $T$ & 0.05 -- 27 & K \\
Pump wavelength &  & 1548.0 & nm \\ 
Acoustic frequency & $\omega/2\pi$ & 10.60 -- 10.62 & GHz \\
Acoustic linewidth & $2 \gamma / 2 \pi$ & 2.1 -- 27.7 & MHz \\
Pump pulse peak power & & 1 & mW \\ 
Pump pulse length & & 5 & $\mu$s \\
Pulse repetition rate & & 500 & s$^{-1}$ \\
Pulse duty cycle & & 0.25 & \% \\
[1ex]
\hline \hline
\end{tabular}
\label{tab:exp_params}
\end{table}

\end{document}